\documentclass[aps,twocolumn,showpacs,preprintnumbers,prb,superscriptaddress]{revtex4}

\usepackage{graphicx}
\usepackage{amssymb}

\newcommand{\R}{\mathbf{r}}

\begin{document}
\title{Meta-GGA exchange-correlation functional with a balanced treatment of nonlocality}

\author{Lucian A. Constantin}
\affiliation{Center for Biomolecular Nanotechnologies @UNILE, 
Istituto Italiano di Tecnologia (IIT), Via Barsanti, 73010 Arnesano (LE), Italy}

\author{E. Fabiano}
\affiliation{National Nanotechnology Laboratory (NNL), Istituto Nanoscienze-CNR,
Via per Arnesano 16, 73100 Lecce, Italy}
\author{F. Della Sala}
\affiliation{National Nanotechnology Laboratory (NNL), Istituto Nanoscienze-CNR,
Via per Arnesano 16, 73100 Lecce, Italy}
\affiliation{Center for Biomolecular Nanotechnologies @UNILE, 
Istituto Italiano di Tecnologia (IIT), Via Barsanti, 73010 Arnesano (LE), Italy}

\begin{abstract}
We construct a meta-generalized-gradient approximation which properly balances
the nonlocality contributions to the exchange and correlation at the semilocal level.
This non-empirical functional shows good accuracy 
for a broad palette of 
properties (thermochemistry, structural properties) and systems 
(molecules, metal clusters, surfaces and bulk solids).
The accuracy for several well known problems in 
electronic structure calculations, such as the bending potential of the 
silver trimer and the dimensional crossover of anionic gold clusters,
is also demonstrated. 
The inclusion of empirical dispersion corrections is finally discussed and 
analyzed.
\end{abstract}


\maketitle

\section{Introduction}
The exchange-correlation (XC) energy functional is the key quantity 
in Kohn-Sham density functional theory (DFT) \cite{ks} and is 
subject to intense research \cite{scus}. 
The simplest functionals of practical utility are those based on the
generalized gradient approximation (GGA) which are
constructed using the electron density and its gradient.
At present, there exist GGA functionals that are rather accurate 
for molecules \cite{PBE,APBE,tognetti08,gg1,gg2},
solids \cite{PBEsol,AM05}, interfaces \cite{PBEint}, and even low-dimensional
systems \cite{q2dgga}. However, in general no GGA functional can be simultaneously
accurate for many of these problems \cite{perdew06}, 
due to the extreme simplicity of the GGA level.
Meta-generalized-gradient-approximations (meta-GGAs) 
\cite{TPSS,revTPSS,regTPSS,mggams,VT08,PKZB,zhao08,M11L}
are the most sophisticated semilocal functionals, incorporating important 
exact conditions, and having an improved overall accuracy with respect to the GGA functionals,
with almost the same attractive computational cost.
These functionals use as an additional ingredient to the GGA ones, the 
kinetic energy density $\tau(\R)$, which enters in the expansion of the angle-averaged 
exact exchange hole \cite{Beckeh}, being thus a natural and important tool in the construction of 
XC approximations. 

In this field, recently we proposed
a meta-GGA dynamical correlation functional (named TPSSloc) \cite{TPSSloc}, 
which takes advantage of  an increased
localization scheme for the correlation
energy density. The TPSSloc correlation has the same form as TPSS \cite{TPSS}, 
but uses as an ingredient a localized version of the PBE correlation functional
\cite{PBE} (named PBEloc), having the physically motivated correlation parameter
(atomic units are used throughout)
\begin{equation}
\beta(r_s,t)=\beta_0+a\;t^2 (1-e^{-r_s^2})\ ,
\label{e1}
\end{equation}
with $\beta_0=0.0375$ obtained from the linear response of the 
local density approximation (LDA) \cite{ks}, $a=0.08$ obtained by 
jellium surface analysis (so that PBEloc/TPSSloc can correctly 
describe quantum effects at the edges of electronic systems \cite{KM}),
$t=|\nabla n|/2 k_s\phi  n$ \cite{LM1}, $k_s=(4k_F/\pi)^{1/2}$ being the
Thomas-Fermi screening wave vector ($k_F=(3\pi^2n)^{1/3}$), 
$r_s=[3/(4\pi n)]^{1/3}$ being the local Seitz radius, 
and $\phi=((1+\zeta)^{2/3}+(1-\zeta)^{2/3})/2$ being a spin-scaling factor 
\cite{zeta1,zeta2}. The TPSSloc correlation functional 
is very accurate for the description 
of correlation effects in the Hooke's atom at all confinement regimes, 
has an accurate and realistic short-range correlation hole, 
and for this reason is more compatible with exact exchange than the 
TPPS \cite{TPSS} or revTPSS \cite{revTPSS} correlation functionals.
In a previous work \cite{TPSSloc}, we used the TPSSloc correlation functional 
in combination with the revTPSS exchange, obtaining good results for
systems where the exact XC hole is localized (atomization energies, 
kinetics and bond lengths of small molecules). 
This approach is however of limited general applicability because, since the
TPSSloc correlation hole density is strongly localized around the electron, 
for broad applicability it needs to be paired with
an exchange
functional which performs (at least) as well as revTPPS exchange for
problems displaying a reduced nonlocality, but also properly describes
the missing nonlocality effects. 
Such an exchange functional should account for most of the tail
XC hole effects \cite{cohencr}, and the static correlation
present in electronic systems near the equilibrium, which can be well
captured by semilocal functionals.
We recall that semilocal exchange functionals describe nonlocality effects 
through a proper shape of the enhancement factor,
especially at medium and large values of the reduced gradient for exchange
$s=|\nabla n|/2 k_F n$, also containing static correlation that scales as exchange under the
uniform scaling of the density, which is essential for the description of
electronic systems with more delocalized density \cite{cohen_science}. 

In this article we consider this problem and
we introduce a non-empirical meta-GGA exchange functional with 
balanced nonlocality contributions, that is fully compatible with the 
TPSSloc correlation functional.
The resulting XC functional with balanced localization is named BLOC. 
Because of a better separation and proper balancing of exchange, dynamical
correlation, and long-range XC effects, can achieve a good accuracy
over a broad range of problems, correcting most of the XC TPSSloc limitations.
In addition, to account for dispersion-related problems,
which cannot be described by our localized correlation nor through 
the exchange nonlocality, we include a semiempirical dispersion correction
\cite{dftd3} to the functional, obtaining the BLOC-D3 functional.

\section{Theory}
\label{sec2}

As starting point of our construction we consider the TPSS and revTPSS
exchange functionals, having the form
\begin{equation}
E_x[n]=\int d\R \; n\; \epsilon_x^{LDA} \; F_x\ ,
\label{e2}
\end{equation}
with $\epsilon_x^{LDA}$ the LDA exchange energy per particle \cite{ks} and
$F_x$ the exchange enhancement factor \cite{TPSS}
\begin{equation}
F_x=1+\kappa -\kappa/(1+x/\kappa)\ .
\label{e3}
\end{equation}
The parameter $\kappa=0.804$ is fixed from the Lieb-Oxford bound \cite{LO},
while
\begin{eqnarray}
x & = & \left[\left(\frac{10}{81}+c\frac{z^f}{(1+z^2)^2}\right)s^2+\frac{146}{2025}\tilde{q}_b^2 -\right.\\
\nonumber
&& - \frac{73}{405}\tilde{q}_b \sqrt{\frac{1}{2}(\frac{3}{5}z)^2+\frac{1}{2}s^4}+ \frac{1}{\kappa}(\frac{10}{81})^2 s^4 +\\
\label{e4}
&& + \left.2\sqrt{e}\frac{10}{81}(\frac{3}{5}z)^2 + e\mu s^6\right]/(1+\sqrt{e}s^2)^2\ ,
\end{eqnarray}
is constructed to recover the fourth-order gradient expansion (GE4) 
of the exchange energy \cite{TPSS}, for a slowly-varying density.
Here, $z=\tau^W/\tau$ ($0\leq z\leq 1$) is the meta-GGA ingredient 
\cite{TPSS} that distinguishes the iso-orbital regions (when $z\rightarrow 1$)
and the slowly-varying regime (when $z \rightarrow 0$), with 
$\tau^W=|\nabla n|^2/(8n)$ being the von Weizs\"{a}cker kinetic energy density 
and $\tau$ the Kohn-Sham positive kinetic energy density;
$s$ is the reduced gradient for exchange; and 
$\tilde{q}_b=(9/20)(\alpha -1)/\sqrt{1+0.4 \alpha(\alpha-1)}+2s^2/3$ 
mimics the reduced Laplacian of the density
$q=\nabla n^{2}/(4(3\pi^{2})^{2/3}n^{5/3})$, 
with $\alpha=(\tau-\tau^W)/\tau^{unif}=(5/3)s^2(1/z-1)$.
The parameters $c$ and $e$ were fixed from the constraint that 
the exchange potential should be finite near the 
nucleus (where $z=1$ and $s\approx 0.4$) and by fitting the 
exchange energy of the hydrogen atom (where $z=1$) \cite{TPSS,revTPSS}.
The parameter $\mu$ dictates the behavior of the functional at 
large $s$ ($s\gtrsim 2$), i.e. in valence and tail regions. 
In TPSS $\mu=0.21951$ which represents the 
non-empirical PBE value, whereas in revTPSS $\mu=0.14$ was set from
semi-empirical considerations. 
Finally, the parameter $f$ controls 
the slowly-varying behavior of the meta-GGA: when $f$ is small 
(e.g. $f=2$ in TPSS) the functional recovers GE4 only when $s$ is 
very small ($s\lesssim 0.1$); when $f$ is larger (e.g. $f=3$ in revTPSS) 
the functional recovers GE4 over a broader range of $s$ values
($s\lesssim 0.3$), and thus gains improved accuracy for bulk solids \cite{revTPSS}.

Due to many successful applications of TPSS and revTPSS, the ansatz 
of Eq. (\ref{e4}) has been proved to be very robust in its general form. 
Thus, it provides an ideal starting point for building our new meta-GGA 
functional through selected modifications and improvements. We recall that
this is a common practice in functional development, so for example the 
TPSS meta-GGA was developed on the skeleton of the PKZB meta-GGA precursor 
\cite{PKZB}, whereas revTPSS is just a judicious modification to TPSS, which 
brings new physical ideas.
We also acknowledge the very recent work \cite{mggams} into the direction 
of simplifying Eq. (\ref{e4}) and solving the order of limits problem. 
However, this new development, despite its conceptual importance, does not
appear, at present, to improve considerably for 
practical molecular applications. 
Therefore, it will not be considered in the present work.

We build our exchange functional using Eqs. (\ref{e2}), (\ref{e3}), and (\ref{e4}), 
and we fix the $\mu$ parameter to its non-empirical PBE and TPSS value
of 0.21951. In this way the good TPSS description at large $s$ 
values, i.e. in the outer-valence and tail regions, is recovered. 
Moreover,
imposing the known meta-GGA constraints \cite{TPSS} for the hydrogen atom, we also have 
$c=1.59096$ and $e=1.537$. Note that these parameters
are fixed only by constraints implying $z=1$, and thus they cannot optimize the
functional for the slowly- and moderately- varying density regimes, i.e. for short-
and partially long- ranges effects. 
This behavior is instead controlled by the value of $f$, 
which determines how fast, for $s\rightarrow 0$, GE4 is 
recovered \cite{revTPSS}, and also 
the behavior of the functional in the valence regions where $0.6\lesssim z<1$ and $s$
is moderately large. 
Thus, a proper choice of $f$ is crucial for the compatibility 
between exchange and correlation parts, being able to modulate
both short- and long-range effects.

In our construction we adopt therefore a flexible choice for $f$ and
generalize it to be a linear function of $z$ (i.e. $f(z)=az+b$).
Note that $z$ is always finite ($0\leq z\leq 1$), and it is a good indicator 
for iso-orbital regions (where $z\approx 1$) as well as for 
the slowly-varying regime (where $z\approx 5s^2/3\ll 1$).
We specify the  following properties:

(i) $f(z)\geq 3$ for $z\leq 0.3$; this condition ensures the 
recovery of GE4 over a wide 
range of $s$ values, and thus the functional can be accurate
for bulk solids \cite{revTPSS}, granting a good description of short-range 
effects. Note that $z\leq 0.3$ in most of the bulk solids, 
where the density is slowly-varying.

(ii) $f(z)\leq 2$ for $z\geq 0.6$; this constraint dictates the 
behavior of the functional in rapidly-varying regions,
giving $F_x \geq F_x^{TPSS}$  in valence and tail regions.

(iii) as for the TPSSloc correlation, we take the jellium surfaces as 
reference system. Thus, we require the exchange functional to complement
the TPSSloc correlation in such a way to be in
agreement with diffusion Monte Carlo (DMC) estimates \cite{WHFGG,ISTLS1} of
jellium surface XC energies (i.e. $\sigma_{xc}^{BLOC}\approx\sigma_{xc}^{DMC}$).
We consider the benchmark $\sigma_{xc}^{DMC}$ given by Eq. (5) of Ref. \onlinecite{ISTLS1}.

Conditions (i)-(iii) are satisfied by the simple form 
\begin{equation}
f(z)=4-3.3 z,
\label{e5}
\end{equation}
which completes the construction of the desired exchange meta-GGA functional
(BLOC exchange).
We note that the form chosen for $f(z)$ is just a simple
ansatz which can satisfy conditions (i)-(iii), without any rigorous base in
an ``exact'' theory. Nevertheless, it provides a simple but effective way
to realize the required balance between exchange and correlation,
 and allows to achieve good
accuracy for the XC functional (see later on).

\begin{figure}
\includegraphics[width=\columnwidth]{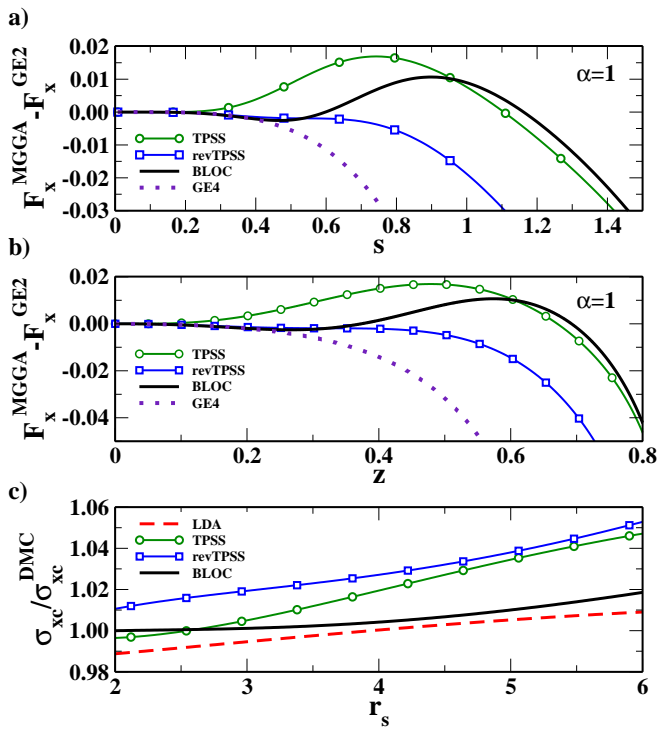}
\caption{Comparison of the exchange enhancement factors of TPSS, revTPSS, GE4, 
and present exchange, in case of $\alpha=1$, with the second order 
gradient expansion $F_x^{GE2}=1+10/81s^2$, versus the reduced gradient $s$ 
(panel a ) and $z$ (panel b).
Panel c shows the ratio of jellium surface XC energies 
$\sigma_{xc}^{approx}/\sigma_{xc}^{DMC}$ versus $r_s$, for several XC functional approximations.} 
\label{fig1}
\end{figure}

The resulting exchange enhancement factor is compared in Fig. \ref{fig1} (panel a and b)
with the TPSS and revTPSS ones, for $\alpha=1$, i.e., in case of 
slowly- and moderately-varying density regimes.
%
%
The plot shows that the new exchange recovers GE4 until $s\leq 0.4$ or
$z\leq 0.3$ (whereas revTPSS recovers GE4 until $s\leq 0.3$), 
being also very close to GE2 until $s\leq 0.6$ ($z\lesssim 0.4$), 
fulfilling condition (i). 
For $s\geq 1$ and $z\geq0.6$ the enhancement factor is 
slightly greater than TPSS, fulfilling condition (ii).

We note that some of the features of the BLOC exchange may recall
the recently proposed VT$\{8,4\}$ meta-GGA exchange functional \cite{VT08}
that is well compatible with revTPSS correlation. Nevertheless, this exchange functional  
slightly de-enhances at $1\leq s \leq 2$ and enhances at $s\geq 2.4$ 
over revTPSS (see Fig. 2 of Ref. \onlinecite{VT08}). Thus, we expect it 
to work similarly as the TPSSloc XC functional \cite{TPSSloc} when used
together with the TPSSloc correlation. 

To complete our analysis, panel c) of Fig. (\ref{fig1}) 
reports the ratio of jellium surface XC energies 
$\sigma_{xc}$ with respect to DMC results  as a function of $r_s$.
These data clearly show that the BLOC XC-functional 
is in best agreement with the DMC benchmark estimates (fulfilling condition (iii)). 
However, we stress that all the functionals presented in the 
figure (TPSS, revTPSS, BLOC, and LDA) are very accurate 
for jellium XC surface energies.

Beyond conditions (i)-(iii), the proposed construction of the BLOC 
exchange carries important physical content and goes beyond the simple 
need to complement the TPSSloc correlation. In fact, it shares most of
the good features of TPSS and revTPSS exchange. Moreover, by construction, 
for any one- and  two-electron systems (where $z=1$), the BLOC exchange 
functional exactly recovers the TPSS one, which is
known to be a very accurate meta-GGA model for exchange in such systems
(being closer to the conventional exact exchange energy density than 
the revTPSS one \cite{revTPSS}).
Thus, BLOC exchange (as all the other non-empirical meta-GGAs) has as a target model 
system also the hydrogen atom 
(note that TPSSloc correlation correctly vanishes for any one-electron system).

\begin{figure}
\includegraphics[width=\columnwidth]{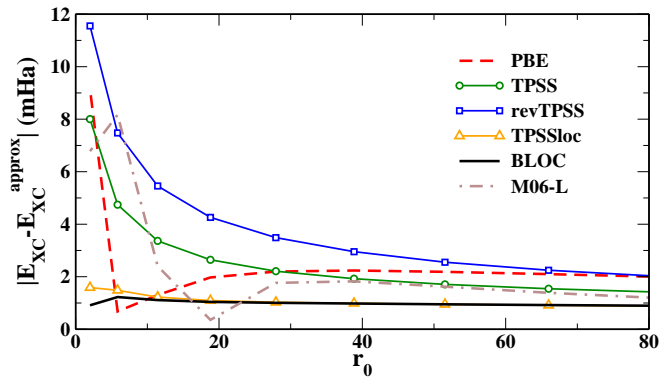}
\caption{Absolute error for the exchange-correlation energy of the Hooke's atom at different values
of the classical electron distances $r_0=(\omega^2/2)^{-1/3}$, with $\omega$ the frequency
of the isotropic harmonic potential. The electrons are tightly bound at small $r_0$ and
are strongly-correlated at large $r_0$.}
\label{fig22}
\end{figure}

Additionally, Fig. (\ref{fig22}) shows that the BLOC XC functional preserves, and
even improves, the performance of the  TPSSloc functional 
for the Hooke's atom, especially in the tightly-bounded region, 
outperforming all the reported meta-GGAs (TPSS, revTPSS, and M06-L \cite{zhao08}). 
We recall that the Hooke's atom \cite{hooke1,hooke2,hooke3,JS,TPSSloc}
represents two interacting electrons in an isotropic harmonic potential.
Thus, a good description of both two-electron exchange and correlation is
required.
This model system provides deep 
physical inside into strongly-correlated and tightly-bounded regimes, being
an important and hard test for density functionals.
%

\begin{figure}
\includegraphics[width=\columnwidth]{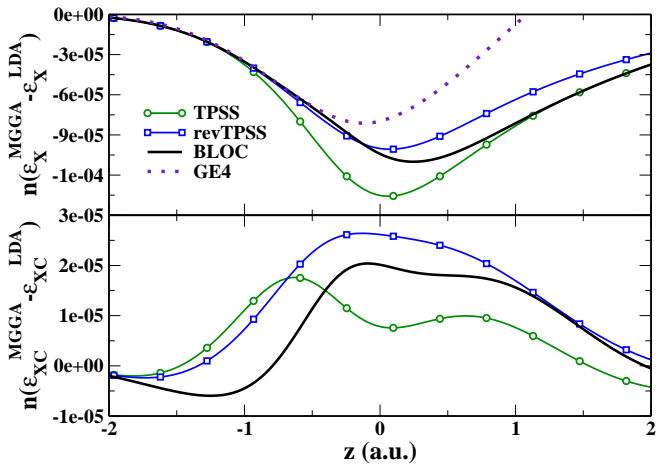}
\caption{Upper panel: GE4, TPSS, revTPSS and BLOC exchange energy densities with respect to LDA 
($n(\epsilon_x-\epsilon_x^{LDA})$), versus the distance $z$, for 
a semi-infinite jellium surface of bulk parameter $r_s=2$. The surface is at $z=0$, 
the bulk is at $z\leq 0$, and the vacuum is at $z\geq 0$.
Lower panel: TPSS, revTPSS and BLOC exchange-correlation energy densities with respect to LDA
($n(\epsilon_{xc}-\epsilon_{xc}^{LDA})$), for the same system.
}
\label{fig23}
\end{figure}

Finally, in Fig. (\ref{fig23}) we show a comparison of the 
TPSS, revTPSS and BLOC exchange (upper panel) and XC (lower panel) energy densities, at 
a jellium surface of bulk parameter $r_s=2$.
The BLOC exchange energy density is smooth and performs in accord with 
conditions (i)-(iii): inside the bulk, where the density is 
slowly-varying, recovers GE4 better than revTPSS, whereas in the vacuum, 
where the density starts to vary rapidly, is close to TPSS.
Note that GE4 fails badly in the tail of density.
On the other hand, the BLOC XC energy density is 
\emph{qualitatively} different from both TPSS and 
revTPSS,
because of the stronger localization of the correlation and of the short-range 
exchange.
%
%
 
\section{Computational details}
\label{sec3}

To assess the performance of the BLOC functional for real problems,
we implemented it into development versions of the
TURBOMOLE \cite{turbomole} and FHI-AIMS \cite{aims1,aims2} program packages.
The two programs were used for all molecular and solid-state calculations, 
using def2-TZVPP \cite{basis1,basis2} and third tier \cite{aims1} basis sets, 
respectively.
All molecular and metal-cluster calculations are self-consistent, 
whereas the bulk solids calculations were performed using PBEsol orbitals and 
densities. A similar non self-consistent approach (using LDA orbitals) 
was used in Refs. \onlinecite{revTPSS,csonka09}. Moreover, we have found that 
the solid-state results are insensitive to the choice of the orbitals (as
also shown in Ref. \onlinecite{csonka09}).
The moderately large basis sets were chosen to provide an
optimal compromise between accuracy and computational cost, so that
our results can be directly transferable to practical applications.
We note also that the use of the same basis set for all the functionals 
in Tab. (\ref{tab1}) should provide a fair assessment of relative performances.
An analysis of the basis set accuracy for the AE6 test is reported
in Supporting Information \cite{supp}.

We considered a set of representative 
and widely used tests (including overall more than 300 systems), which were 
divided into seven groups:
\begin{itemize}
\item \textbf{Atomization energies and proton affinities of organic molecules}, 
which includes the W4 \cite{w4,grimme1,grimme2} and PA12 \cite{grimme1,grimme2}
test sets. Moreover, the AE6 test \cite{ae6} for atomization energies was
also chosen, because this small representative test is often used 
to benchmark quantum chemistry methods and provides therefore a direct 
comparison with many other works. Finally, the W4-MR test \cite{w4},
of molecules with non-single-reference character, was considered
for completeness.
\item \textbf{Reaction energies and kinetics}, including representative tests
for organic reaction energies (OMRE \cite{omre}), barrier heights
(BH6 \cite{ae6}) and both (K9 \cite{k9}).
\item \textbf{Other} tests, which are relevant for assessing density functionals: 
ionization potentials (IP13 \cite{lynch03}), isomerization energies (ISOL6 \cite{luo11}), 
difficult cases for DFT (DC9/12 \cite{peverati12}), and 
absolute atomic energies (AE17 \cite{zhao08}).
\item \textbf{Non-covalent interactions}, comprising tests for hydrogen bonds 
(HB6 \cite{hb6}), dipole-dipole interactions (DI6 \cite{hb6}), 
$\pi-\pi$ stacking (PPS5 \cite{hb6}), and the S22 test \cite{grimme1,grimme2},
which includes a broad selection of non-covalent interactions.
\item \textbf{Structural properties} of organic molecules. This group includes
a test of optimized bond lengths (MGBL19 \cite{mgbl19}) and one
test of harmonic vibrational frequencies (F38 \cite{f38}).
\item \textbf{Transition metals}. In this group we collected a set of tests 
involving transition metal complexes (TM10AE \cite{zeta1} for atomization
energies and TMBL \cite{buhl06} for bond lengths) and gold clusters
(AunAE \cite{gold,zeta1} for atomization energies and AuBL6 \cite{gold}
for bond lengths).
\item \textbf{Bulk solids}. This group includes the tests of the
equilibrium lattice constants (LC12), bulk moduli (BM12), and
cohesive energies (COH12) of 12 bulk solids: Li, Na, Al (simple metals);
Cu, Ag, Pd (transition metals); Si, Ge, GaAs (semiconductors);
NaCl, NaF, MgO (ionic solids). The reference data were taken from
Ref. \onlinecite{csonka09}. 
\end{itemize}

To assess the performance of different functionals for each group of
tests we consider, beside the individual mean absolute error (MAE) for
each test, an overall MAE relative to TPSS (RMAE), defined as \cite{mukappa}
\begin{equation}
\mathrm{RMAE} = \frac{1}{M}\sum_i^M\frac{\mathrm{MAE}_i}{\mathrm{MAE}_i^\mathrm{TPSS}}\ ,
\end{equation}
where the sum runs over all tests (M) within a group and
$\mathrm{MAE}_i^\mathrm{TPSS}$ is the MAE of TPSS for the $i$-th test.
The RMAE provides an indication of whether any method is 
better (RMAE$<1$) or worse (RMAE$>1$) than TPSS for a given problem. 
Thus, it allows a fair global assessment of all the results.

Finally, in addition to the tests listed above we considered two special
cases which are known to be difficult for semilocal density functionals,
namely the description of the bending potential of the silver trimer \cite{sil1,sil2}
and the dimensional crossover of gold anionic clusters \cite{johansson08,furche02,hakkinen03,xing06}.

\section{Results}
The results of all tests are summarized in Tab. \ref{tab1} where
we report the MAEs and RMAEs for different tests and groups
as resulting from different functionals. The TPSS and revTPSS
results are considered for direct comparison with parent non-empirical functionals.
The M06-L results are
also reported, to have a reference of a widely
used and highly parameterized meta-GGA functional.
\begin{table*}[hbt]
\begin{center}
\caption{\label{tab1} Mean absolute errors for various tests as resulting from TPSS, revTPSS, TPSSloc,
BLOC, and BLOC-D3 meta-GGAs calculations. M06-L results are also reported for reference.
The BLOC results are underlined whenever they equate or are better than both TPSS and revTPSS.
For each group of tests the mean absolute errors relative to TPSS (RMAEs) are reported in the last line. 
Full results are reported in Supporting Information \cite{supp}.}
\begin{ruledtabular}
\begin{tabular}{lcccccc}
Test set & TPSS & revTPSS & TPSSloc & BLOC & BLOC-D3 & M06-L  \\
\hline
\multicolumn{7}{c}{Atomization energies and proton affinities (kcal/mol)}\\
organic molecules (AE6)          & 5.4 & 6.6 & 3.9 & \underline{3.6} & 3.6 & 3.4  \\ 
organic molecules (W4-NMR)       & 4.7 & 5.2 & 5.5 & \underline{4.6} & 4.4 & 5.4 \\
static correlation (W4-MR)       & 8.8 & 9.4 & 14.0 & 8.9 & 9.1& 5.9 \\ 
proton affinities (PA12)          & 4.7 & 4.8 & 3.8 & \underline{3.7} & 3.8 & 4.5 \\   
RMAE                             & {\bf 1.00} & {\bf 1.10} & {\bf 1.07} & \underline{{\bf 0.86}} & {\bf 0.86} & \textbf{0.85} \\
\multicolumn{7}{c}{Reaction energies and kinetics (kcal/mol)}\\
organic molecules (OMRE)         & 8.0 & 10.2& 7.9 & \underline{7.1} & 6.0 & 5.3 \\  
kinetics (K9)                    & 7.0 & 7.2 & 6.5 & \underline{6.3} & 6.4 & 4.2 \\
barrier height (BH6)             & 8.3 & 7.4 & 8.6 & 8.9 & 8.9 & 4.3 \\              
RMAE                             & {\bf 1.00} & {\bf 1.07} & {\bf 0.98} & \underline{{\bf 0.95}} & {\bf 0.91} & \textbf{0.59} \\
\multicolumn{7}{c}{Other (kcal/mol)}\\
ionization pot. (IP13)           & 3.1 & 2.9 & 3.0 & 3.3 & 3.3 & 3.1 \\         
isomerization (ISOL6)            & 3.6 & 4.0 & 2.7 & \underline{3.5} & 3.0 & 2.8 \\
difficult cases (DC9/12)         & 18.2 & 21.7 & 29.6 & 20.1 &  19.4 & 22.3 \\
atomic energies (AE17)           & 22.6 & 41.8 & 42.4 & 26.8 & 26.8 & 3.9 \\
RMAE                             & {\bf 1.00} & {\bf 1.27} & {\bf 1.31} & {\bf 1.08} & {\bf 1.04} & \textbf{0.79} \\
\multicolumn{7}{c}{Non-covalent interactions (kcal/mol)}\\
hydrogen bonding (HB6)           & 0.6 & 0.6 & 0.6 & \underline{0.6} & 0.8 & 0.6 \\ 
dipole bonding (DI6)             & 0.6 & 0.5 & 0.5 & \underline{0.5} & 0.8 & 0.4 \\
$\pi-\pi$ stacking (PPS5)        & 2.8 & 2.5 & 2.7 & 2.8 & 0.1 & 0.8 \\
various non-covalent (S22)       & 3.2 & 2.8 & 2.6 & 3.3 & 0.5 & 0.6 \\
RMAE                             & {\bf 1.00} & {\bf 0.90} & {\bf 0.90} & {\bf 0.97} & {\bf 0.71} & \textbf{0.53} \\
\multicolumn{7}{c}{Structural properties (m\AA{} and cm$^{-1}$)}\\
bond lengths (MGBL19)            & 6.9 & 7.4 & 6.8 & \underline{5.9} & 5.7 & 3.0 \\         
vibrations (F38)                 & 44.1 & 43.7 & 41.9 & \underline{39.0} & 40.0 & 41.5 \\
RMAE                             & {\bf 1.00} & {\bf 1.03} & {\bf 0.97}  & \underline{{\bf 0.87}} & {\bf 0.87} & \textbf{0.69} \\
\multicolumn{7}{c}{Transition metals (m\AA{}, kcal/mol (TM10AE), and kcal/mol/atom (AUnAE)}\\
transition-metal energies (TM10AE) & 10.8 & 11.1 & 10.3 & \underline{11.1} & 11.0 & 7.9 \\
gold cluster energies (AUnAE   )   & 0.6  & 1.6  & 4.4  & \underline{0.3}  & 0.4 & 1.4 \\
transition-metal geom. (TMBL)      & 12.7 & 11.6 & 17.7 & \underline{11.2} & 11.3 & 10.9 \\
gold cluster geom. (AuBL6)         & 30.8 & 21.9 & 35.3 & 23.3 & 19.6 & 33.9 \\
RMAE                               & {\bf 1.00} & {\bf 1.33} & {\bf 2.71} & \underline{{\bf 0.79}} & {\bf 0.80} & \textbf{1.26} \\

\multicolumn{7}{c}{Global assessment for chemical properties} \\
Average RMAE                     & {\bf 1.00} & {\bf 1.12} & {\bf 1.32} & \underline{{\bf 0.92}} & {\bf 0.87} & \textbf{0.79} \\
\hline
\multicolumn{7}{c}{Bulk solids (\AA{}, GPa, and eV/atom)}\\
lattice constants (LC12)          & 0.056 & 0.043 & 0.049 & 0.048 & - & 0.057 \\
bulk moduli (BM12)                & 7.8 & 9.4 & 15.7 & \underline{7.8} & - & 10.5 \\
cohesive energies (COH12)         & 0.131 & 0.162 & 0.385 & \underline{0.117} & - & 0.358 \\
RMAE                             & {\bf 1.00}  & {\bf 1.07}  & {\bf 1.94} & \underline{{\bf 0.92}} & - & \textbf{1.70} \\
\end{tabular}
\end{ruledtabular}
\end{center}
\end{table*}

Inspection of the Tab. (\ref{tab1}) shows that BLOC performs better than both revTPSS and 
TPSS in more than $60\%$ of the tests.
On the other hand, it compares also relatively well with the ``benchmark'' M06-L
results, yielding lower MAEs in 30\% of the cases and giving significantly
worse results only in two cases (atomic energies and barrier heights).
Even though this performance 
may not appear to be outstanding at first sight, we remind the reader of 
the difficulty of outperforming
highly parameterized meta-GGA functionals without resorting to
higher non-empirical rungs of the DFT Jacob's ladder \cite{ladder}.
At the same time, the comparison of BLOC, TPSS, and revTPSS 
results shows how difficult it is to reach a good accuracy at meta-GGA level,
for a broad range of properties, without introducing a high level of
empiricism. In this sense, the good performance of BLOC is quite
satisfactory, and its rather systematic improvement over TPSS and revTPSS 
indicates that the functional captures well the essential physics of a wide selection of 
electronic systems at the simple semilocal level. Thus, the balanced 
description of different regimes and the physical concepts included in exchange 
and correlation parts of the BLOC functional are relevant.

In particular, we underline here the 
good performance of the BLOC functional for the structural 
properties of molecules, 
atomization energies/proton affinities, 
and all the properties of transition metal systems.
On the other hand, in only three cases (BH6, IP13, S22) BLOC is worse than both
revTPSS and TPSS (still being very close to the TPSS). Moreover,
at least two of these tests (BH6, S22) 
can hardly be accurate for non-empirical semilocal functionals.
We also remark that HF+TPSSloc is extremely good for BH6 (see Table I of Ref. \onlinecite{TPSSloc}),
and thus we expect that a hybrid of BLOC functional can be rather accurate for barrier heights.

Overall BLOC has the second best RMAE (0.92) for chemical 
properties and the best RMAE for bulk solids (0.92).
Note that the M06-L functional is the best for molecular properties,
but it is modest for solids. The more recent M11-L functional \cite{M11L}
improves lattice constants, but not solid-state cohesive energies \cite{NEW1}.
In fact, even revTPSS improves over TPSS for lattice constants, but  
worsens both bulk moduli and cohesive energies (see Tab. \ref{tab1} and also Ref. \onlinecite{NEW2}).
On the other hand, BLOC gives a very good balance for all solid-state properties.

We remark also that notably these results can be in general obtained with a
small error compensation 
for situations where the XC hole is reasonably localized (e.g. for the AE6 test), 
because of the higher compatibility of the TPSSloc correlation with 
exact exchange \cite{TPSSloc}. In these cases in fact TPSSloc and
BLOC correctly yield very close results. On the other hand, for
systems characterized by delocalized electrons and/or static correlation,
our balanced exchange functional proves to be capable to compensate 
well for the missing long-tail behavior of the TPSSloc correlation,
so that good results can be finally obtained using the BLOC XC functional
(see e.g. W4-MR, DC9/12, AunAE, and COH12).

The results of Tab. \ref{tab1} show (second last column) that a small further improvement 
can be achieved by complementing the BLOC functional with semiempirical 
dispersion corrections. In fact, dispersion cannot be described 
by the localized TPSSloc correlation nor can it be included in the exchange
part. We implemented this correction through the
DFT-D3 semiempirical model \cite{dftd3}, fixing the two free parameters
of the model by fitting to the MAE of the S22 test. The resulting parameters are
$s_{r,6}=1.104$ and $s_8=0.888$, slightly smaller than the TPSS values.
We see that the BLOC-D3 functional of course strongly improves the 
performance for dispersion-dominated tests (e.g., S22, PPS5) but, 
at the same time, also preserves the accuracy for all tests, 
even slightly improving the MAE in some cases (e.g. structural properties, 
W4, OMRE). Thus, the semiempirical DFT-D3 correction appears to 
integrate well with our construction,
and as a result the
BLOC-D3 functional yields a very good performance for chemistry-related
tests (RMAE=0.87).

%
\begin{figure}
\includegraphics[width=\columnwidth]{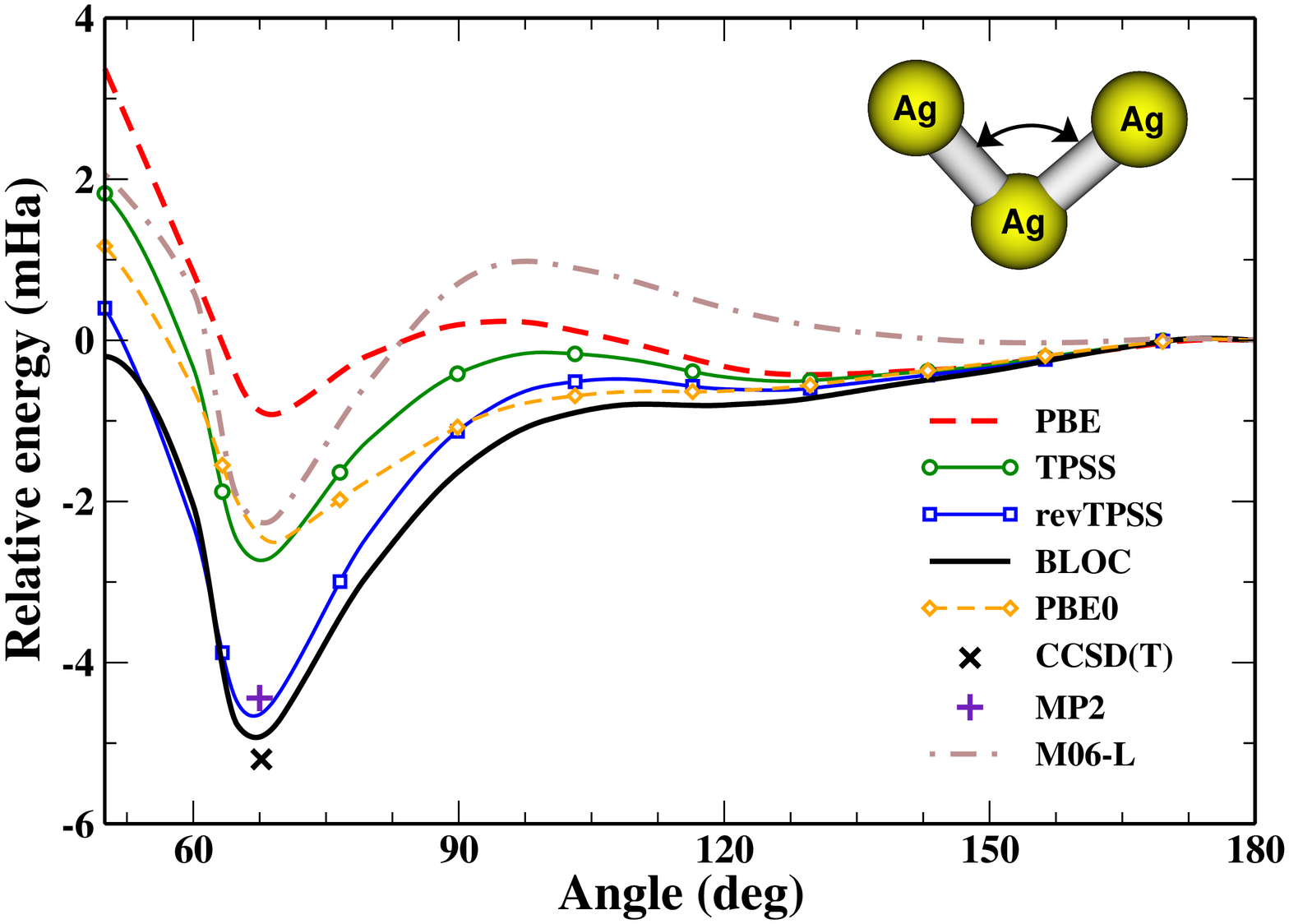}
\caption{Relative energy (in mHa) with respect to the linear structure 
of Ag$_3$ ($\Delta E=E(\theta)-E(180^\circ)$) as a function of 
the Ag-Ag-Ag bond angle $\theta$ (in degrees) for several functionals. 
CCSD(T) and MP2 values are taken from Ref. \onlinecite{sil2}.}
\label{fig2}
\end{figure}
%

\subsection{Silver trimer}
\label{sub1} 
As additional test for our assessment of the BLOC functional 
we consider a particularly difficult case for DFT: the description of the
potential energy related to the bending of the Ag$_3$ cluster \cite{sil1,sil2}.
For this case in fact high-level methods indicate the existence of a single
minimum at a bond angle of $\theta\approx 70^\circ$, whereas all PBE-based 
functionals yield two almost degenerate minima at $\theta\approx 70^\circ$ and 
at $\theta\approx 140^\circ$ \cite{sil1}, and all LYP-based functionals 
yield only the wrong $140^\circ$ minimum. 
The results reported in Fig. \ref{fig2} show indeed that the popular PBE
and even the meta-GGA TPSS or the hybrid PBE0 functionals yield a
poor quantitative description of the problem, where the relative energy of the
global minimum is underestimated, and only by the
inclusion of exact exchange 
can the second minimum at $\theta\approx 140^\circ$ be turned into a shoulder.
On the other hand, revTPPS yields a good relative
energy for the minimum at $\theta\approx 70^\circ$, but still displays a
small second minimum at $\theta\approx 140^\circ$, due to a hill at $\approx 100^\circ$.  
The BLOC functional instead gives a very good description of 
the bending potential of the Ag$_3$ cluster
in good agreement with CCSD(T) calculations \cite{sil2}, transforming the 
hill at $\approx 100^\circ$ into a shoulder.
We note finally, that for this problem highly empirical meta-GGAs (as M06-L) are 
also rather inaccurate (as shown in Fig. \ref{fig2}).

\subsection{The dimensional transition of anionic gold clusters problem}
\label{sub2}
Finally, we consider the problem of
the 2-dimensional to 3-dimensional (2D$\rightarrow$3D)
transition in gold cluster anions which attracted great interest
over the last several years
\cite{johansson08,furche02,hakkinen03,xing06,mantina09,ferrighi09}.
A joint electron diffraction and DFT study
\cite{johansson08} showed in fact that the 2D$\rightarrow$3D transition
occurs for Au$_n^-$ clusters at $n=12$, where a 2D and a 3D
structure are almost isoenergetic, while planar (2D)
and 3D structures characterize the experimental spectrum for $n=11$ and
$n=13$, respectively. However, most DFT functionals are unable to reproduce 
this outcome \cite{johansson08,mantina09,ferrighi09}.
%
\begin{table}
\begin{center}
\caption{\label{tab2} Relative energies (meV) with respect to conformer I of 2D and 3D gold
cluster anions with 11-13 atoms. The structures of the clusters and the nomenclature can be found
in Ref. \onlinecite{johansson08}. All energies include the correction terms
reported in Table IV of Ref. \onlinecite{johansson08}, in order to account
for spin-orbit, all-electron, zero-point
vibrational-energy, and thermal effects.}
\begin{ruledtabular}
\begin{tabular}{llrrrrr}
System & dim$^a$ &  \multicolumn{1}{c}{TPSS} & \multicolumn{1}{c}{revTPSS}
& \multicolumn{1}{c}{TPSSloc} & \multicolumn{1}{c}{BLOC} & \multicolumn{1}{c}{M06-L$^b$} \\
\hline
Au$_{11}^-$-I  & 2D          &   0 &   0 &   0 &   0  & 0 \\
Au$_{11}^-$-II & 3D          & 190    &  159 & 121 & 190 & 60 \\
Au$_{11}^-$-III& 3D          & 270    &  337 & 419 & 310 & 250 \\
               &             &        &        &        &     &   \\
Au$_{12}^-$-I  & 3D          &   0 &   0 &   0 &   0 & 0 \\
Au$_{12}^-$-II & 2D          & -170   & 79 & 343 & 37 & 400 \\
               &             &        &        &        &        \\
Au$_{13}^-$-I  & 3D          &   0 &   0 &   0 &   0 & 0 \\
Au$_{13}^-$-II & 3D          & 10     &   39 &  65 &  29 & 110 \\
Au$_{13}^-$-III& 2D          & 230    & 467 & 766 & 426 & 930 \\
\end{tabular}
\end{ruledtabular}
\end{center}
\begin{flushleft}
a) Dimensionality.\\
b) Data from Ref. \onlinecite{mantina09}
\end{flushleft}
\end{table}
%

In Tab. \ref{tab2} we show the relative energies of several 2D and 3D gold clusters  
as obtained from several meta-GGA functionals.
The TPSS functional erroneously predicts the 2D structure to be favorable at $n=12$
(as most other GGA functionals, including PBE \cite{PBE,gold}). 
The other meta-GGAs (revTPSS, TPSSloc, BLOC, and M06-L), on the other hand,
correctly solve the dimensional crossover of anionic gold clusters.
However, only the revTPSS and BLOC meta-GGAs predict almost
isoenergetic structures for Au$_{12}^-$-I and Au$_{12}^-$-II,
in better agreement with experiment \cite{johansson08}, and with the M06 hybrid functional 
\cite{mantina09}.

\section{Conclusions}
In conclusion, we constructed an accurate, non-empirical and physically motivated
meta-GGA XC functional.
Its dynamical correlation part is strongly localized, showing a realistic short-range correlation, 
and being more compatible with exact exchange than other popular 
correlation functionals \cite{TPSSloc}. The exchange part was constructed
to balance nonlocality effects at the semilocal exchange level,
so to have a good description of different density regimes in different systems.
Additional inclusion of dispersion effects can be efficiently obtained 
via semiempirical corrections. 
 
Nowadays, highly-empirical functionals are optimized against a 
large set of data and chemical properties 
\cite{zhao08}, and thus they 
become very attractive for many electronic structure calculations. 
However, due to their empirical nature, they may show unexpected
failures for specific or exotic problems.
On the other hand, the 
non-empirical functionals, constructed from 
exact quantum mechanics conditions and from model systems, show a reasonable accuracy for 
most of applications, due to the physics which they incorporate.
The BLOC meta-GGA improves over, or is in line with, state-of-the-art TPSS and 
revTPSS for many 
energetic and structural properties 
of organic molecules, transition metal complexes and clusters, jellium surfaces, Hooke's atom,
 and bulk solids,  can 
correctly solve difficult problems 
as bending potential of the silver trimer and the dimensional crossover in gold anionic clusters,  
and thus can become a non-empirical workhorse tool in quantum 
chemistry and condensed matter.

\section{acknowledgement}
This work was partially funded by the European Research Council (ERC) 
Starting Grant FP7 Project DEDOM, Grant No. 207441. We thank 
TURBOMOLE GmbH for providing the TURBOMOLE program package and M. 
Margarito for technical support.

\end{document}